\documentstyle[12pt]{article}
\title  {Invariant quantization  in  \\
 warped spacetimes}

\medskip

\author { Philippe Droz-Vincent\\[2mm]
 LUTH\\
5 place Jules  Janssen, Meudon,  France}
\date {\    }
\newcommand {\batonR}{ \mbox{I\hspace*{-1mm}R} }

\newcommand{\sig}{\sigma}

\newcommand{\AB}{ {AB} }  

\newcommand{\beq}{\begin{equation}}
\newcommand{\eeq}{\end{equation}}
  \newcommand{\dron}{\partial}   
 \newcommand{\beprop}{\begin{prop}}
 \newtheorem{prop}{Proposition}    
 \newcommand{\enprop}{\end{prop}}
   \newtheorem{theo}{Theorem}   
   \newcommand{\betheo}{\begin{theo}}
   \newcommand{\entheo}{\end{theo}}
\newcommand{\alp}{\alpha}             
\newcommand{\gam}{\gamma}
\newcommand{\Gam}{\Gamma}
\newcommand {\half}{ {1 \over 2}} 
\newcommand {\noi}{\noindent}

\newcommand {\lam}{\lambda}
\newcommand{\calk}{ {\cal K} }
\newcommand  {\disp}{\displaystyle}

\newcommand{\del}{\delta}     
\newcommand {\Del}{\Delta} 
\newcommand {\Sig}{\Sigma}

\newcommand {\fhat}{\widehat f} 
\newcommand \hhat {\widehat h}
\newcommand {\nab}{\nabla}

\newcommand{\aron}{{\cal A}}
\newcommand{\eron}{{\cal E}}
\newcommand{\sron}{{\cal S}}
\newcommand{\hron}{{\cal H}}    
\newcommand{\fron}{{\cal F}}   
\newcommand{\cron}{{\cal C}}
\begin{document}
\maketitle
\abstract{ We argue that  quantum theory in curved spacetime  should 
be invariant under the continuous spacetime symmetries that are connected with 
the identity. 
For the most typical  warped-product spacetimes,  we prove that 
such invariance can be actually implemented, at least at the level of first 
quantization. Indeed, in the framework of warped spacetimes,
  we can ensure the isometric invariance of the projection operator that 
 selects   the positive-energy piece of any solution to the wave equation.
 Of course, for a given spacetime,  this operator is not unique,
 but  the requirement of invariance under spacetime isometries drastically
 reduces the arbitrariness.

\noi
Quantum theory of free fields additionally requires that this  linear operator
admits a  kernel. 
We briefly discuss conditions implying that an  invariant kernel
 actually exists, and we characterize the general form it  can take on.

\section {Introduction}

There are two ways for constructing a quantum theory of free fields.

\noi The traditional formulation starts with a single-particle Hilbert space, 
then considers Fock space, and further introduces field operators through 
creation and anihilation operators associated with the 
so-called "two-point function". 

\noi In contrast, the algebraic formulation, immediately starts from a
 $C^*$-algebra of fundamental  observables; then, it recovers the concept of
 state vector with help of  sophisticated mathematical developments.

\noi From a very general point of view, it is true that the algebraic approach 
is more powerful: in many cases, the various possible quantizations permitted
 by the traditional formulation are not all  unitarily
   equivalent  among  themselves,
whereas  the inequivalent representations  correspond to
$  C^*$-algebras of fundamental observables that are, however, isomorphic.

\noi
   Nevertheless, in this article, we shall stick  to the traditional 
approach,
first because of a personal preference for the most intuitive setting,
and  also because we keep in mind   that, for a large class of spacetimes
 (for instance, in the case of a spatially closed universe, 
see  \cite{wal} pp. 96-97)
the traditional formulation of quantum field theory remains essentially as
 powerful as the algebraic approach.

We  naturally start with "first quantization", and consider the wave equation 
for a $c$-number wave function. Indeed,   in spite of  
well-known limitations concerning the physical interpretation of one-body
 relativistic quantum mechanics,   
it is of conceptual interest to investigate how far one can go without 
inconsistencies by developping a single-particle quantum theory for its 
own right.
 And even from the viewpoint of quantum field theory, a  careful construction
 of the  one-particle sector is a technical prerequisite.                   

In fact, in the traditional approach, it is easy to disentangle  
first quantization from the issues that are genuinely concerned  with 
quantum field theory; this way of proceeding allows to avoid unnecessary
 complications.
 This remark typically applies to the problem of  separating
 positive from negative-energy solutions.

\noi
So the first step consists in setting up the Hilbert space for a single 
particle;  the construction of Fock space will  be performed ultimately, 
with help of a  positive-energy kernel associated with the projection operator.

\medskip
Insofar as {\em scalar}  particles are concerned,
 one starts from   the Klein-Gordon (KG) equation.
 Its solutions  form a linear space $\calk$ endowed with a natural sesquilinear 
form; this form is by no means positive.
But a positive definite scalar product (and ultimately a  Hilbert space)
 will be obtained further, 
 provided  one is able to perform a convenient  splitting 
of $\calk $  into  positive and negative-frequency solutions.

This splitting corresponds to the action of a  projection linear operator
   $\Pi ^+$,           or equivalently by that of a so-called 
 {\em complex-structure operator} $J$ which 
must enjoy a special algebraic property with respect to the sesquilinear form
 (in fact $J$ must be, in certain sense, {\em positive} with respect to  
 the underlaying skew-symmetric form $w$).

\medskip 
\noi        In as much  as  linear operators acting in $\calk$
 can be represented by bi-scalar kernels, an alternative  formulation of 
the splitting can also be made in terms  of  a positive-frequency 
kernel  $D^+$ which is sometimes referred to as the
 two-point Wightman "function".
In contradistinction to other propagators and Green functions, in general 
this kernel {\em is not}  uniquely provided by the geometry.
 Since  all the properties of the quantum  theory of free fields are encoded
 in it, the knowledge of this kernel is  usually considered as providing 
a definition of the vacuum. 

\medskip
\noi
For an arbitrary given spacetime, the existence of a splitting which separates 
positive-frequency solutions from negative-frequency solutions is not
 problematic,  but  its lack of unicity has motivated an abundant 
literature. First results where obtained about static and stationary 
spacetimes \cite{cheval} \cite{mor};
 apart from these special cases, there is generally 
no preferred splitting. 
Soon it was realized that, under very general assumptions including  global 
hyperbolicity,  each spacelike Cauchy surface  allows for a 
different  definition of the frequency splitting \cite{ashmag}.

\noi    This result was cast into a rigorous form by C. Moreno \cite{moreno}
who proved the symplectic (resp. unitary) equivalence of these different 
definitions; he took a   step further by directly  constructing  the complex
 structures  in terms of their  kernels, according to the Lichnerowicz 
program \cite{lichhouch}. 

Thus, insofar as one ignores the issue of symmetries, the construction  of 
quantum mechanics in a general spacetime 
 is, to a very large extent, a solved problem.

 \medskip
 \noi
But, in the framework of General Relativity, it is natural to demand that
 quantum theory be invariant under spacetime  isometries. 
We have elsewhere \cite{eres} proposed  a  principle of isometric invariance:

\noi   {\sl Quantum mechanics of free particles must be invariant under 
all spacetime isometries continuously connected with the Identity}

\noi
This condition  is a natural generalization of the requirement of  
  Poincar\'e invariance  which is at the very foundation of quantum field 
theory in Minkowski spacetime.  
Here, spacetime symmetries  generate first integrals for the motion of a test 
particle (this is true at least for minimal coupling). Insofar as these 
constants of the motion have a physical meaning, they should be represented 
by operators acting in the positive-frequency sector (in the hope of 
ultimately being  implemented as  essentially self-adjoint operators).
 They must therefore map this sector into itself,
 which  requires that they commute with $\Pi ^+$.
Thus, in a spacetime admitting  Killing vectors, imposing isometric 
invariance doesnot remove but substancially  restricts the arbitrariness
in the choice of $\Pi ^+$.   For a given spacetime of this kind, the first 
question, of course, is whether an  invariant quantization is possible at all.
At first sight, it may be possible to construct an invariant frequency 
splitting by using an invariant Cauchy surface. This procedure is obviously 
limited to cases where all Killing  vectors are spacelike, and faces the 
question as to determine a Cauchy surface which is invariant under the action 
of all  of them.
    
\medskip
\noi
Note that the invariance of $\Pi ^+$ may   be ensured by that of  its kernel
  $D^+$.
But,  whereas   the isometric invariance of  retarded and advanced 
Green functions  has been  known for a long time \cite{lichne},  
most available results ensuring the existence of  $D^+$ in a large class of
 spacetimes tell practically nothing  about the possible invariance of 
this  kernel  under isometries. 
There are a few exceptions: the issue of constructing an invariant vacuum 
has been discussed for de Sitter spacetime \cite{sit}; see  a remark in Birell 
and Davies \cite{bir} about invariance of the Wightman "function" in an 
asymptotically static, spatially flat FRW universe.
 In a  previous work \cite{eres} we presented the case of generalized FRW 
universes.

\medskip

\noi In this article, we focus on warped product manifolds because they have 
the following properties:  

First, their isometry  group is, to a large extent, under control,
 owing to the works of  Carot and da Costa \cite{carot} and 
M. S\'anchez \cite{sanch}.  

Second, the wave equation in such spacetimes admits a remarkable constant of 
the motion,   which allows for considering sectors of  
 {\em mode solutions}. 
This situation leads to constructing an invariant energy-splitting separately 
for each mode.
In fact, the wave equation undergoes a separation of the variables and one is 
left with a reduced equation involving a minor number of degrees of freedom.

\medskip \noi
This paper is organized as follows:
 Section 2 contains basic formulas, a display of the notation and terminology.
Warped spacetimes and known results about 
their isometries are recalled in Section 3. In Section 4 we perform the
  separation of the variables according to the notion of modes and 
reformulate  our primitive  problem in a reduced manner,
 taking advantage of the fact that each mode is a tensor-product.
Section 5 is devoted to  general facts of bilinear algebra about skew forms,
 sesquilinear forms and complex structures in a tensor product.
  In Section 6 we characterize a  "canonical" form of the splitting we 
were looking for, and perform a sum over modes in order to construct the 
sector of positive-frequency solutions. 

\noi Most part of this article is devoted to the single-particle problem.
We are thus concerned with the KG equation for a c-number field.
Next step would  be the construction (in invariant manner) of creation and 
anihilation operators in Fock space.
This program requires that our  splitting actually admits a kernel, 
 and that this kernel in turn respects isometric invariance.
Kernels and  two-point functions are briefly discussed in Section 7.
The final  Section is about remarks and conclusions.
 Appendix 1 deals with the KG equation with a source term, 
Appendix 2 offers a summary of useful elementary calculations.

\section{Basic formulas}
\noi
Spacetime is considered as given, and we are not concerned with 
stress-energy tensors.
Throughout this paper we consider smooth metrics, smooth  functions, 
 and linear operators acting in various functional spaces.
Our point of view  is that of differential geometry; operators  
and eigenvalues are understood  in the sense of the geometric spectral 
theory \cite{geomspec}.

\noi  As long as possible, we remain within the framework 
of multilinear algebra and postpone several issues  involving the continuity of 
the operators.                                                
But, of course, continuity becomes  essential as soon as one whishes
 to associate 
$\Pi ^+$ with a kernel $D^+ (x,y)$ (a distribution), as is necessary
 in order to consider Fock space.
                          
 The main tool for  connecting  with topologic  matters is the 
observation that, the operators we  construct with help of Cauchy surfaces are 
actually continuous because the functions determined by their Cauchy data 
depend continuously on them. 

\medskip \noi
We consider here the Klein-Gordon equation 
\beq   (\nabla ^2 + m^2 ) \psi = 0               \label{weq}    \eeq
for a   wave function $\psi (x)$  describing the minimal coupling 
of a scalar particle with gravity.
Complex (resp. real) solutions to (\ref{weq}) live in an  infinite dimensional 
vector space  $\calk$    (resp.  $ \calk ^R $  ).
The   real  and   skew-symetric   form                          
\beq       \varpi (\phi , \psi ) = 
 \int _\Sig   (\phi \nab ^\mu \psi -  \psi  \nab ^\mu \phi )  
 \       d \Sig _\mu            \label{defw}    \eeq
is  conservative with respect to changes of hypersurface $\Sigma$,  provided 
$\phi $ and $\psi$ are solutions to the KG equation.                         
Under very general assumptions, $\varpi$ is in fact a {\em symplectic form}.

\noi For complex  $\phi$  and $\psi$, the  sesquilinear form 
\beq    (\phi ; \psi) =    -i  \varpi  (\phi ^*  , \psi ) =
     \int j^\mu (\phi, \psi)  \   d\Sigma _ \mu             \label{sesq} \eeq
 {\em is not} positive definite.
       Thus, in order to exhibit a candidate for one-particle pre-Hilbert 
space,   the linear space of solutions $\calk$  
must be split as the direct sum of  two subspaces.
  In one of them (further completed and  identified as the positive-frequency
 sector) the restriction of $(\phi ; \psi)$  must be definite  positive.
\noi The trouble is that  such   splitting is  not unique   and,  
except for the case  of  stationary spacetimes  \cite{cheval} \cite{mor} 
 there is no natural  way to   select a preferred choice.
 Nevertheless, criteria for the   determination of this splitting have been
 given soon,  either in terms of finding a real linear
 operator $J$ with $J^2 = - 1$, determining a complex structure in the space
 of real solutions ~\cite{segal},   
or equivalently  in terms of a  projector $\Pi ^+ = \half (1 + i J)$
which projects any complex solution into the positive-frequency subspace.

\noi  In fact, $J$ must be {\em positive}
 with respect to the skew-symmetric form $\varpi$.
This property is more restrictive than simply leaving the skew form  invariant;
  it means both following  conditions
\beq   \varpi ( J \phi , J  \psi )  = 
  \varpi (\phi , \psi ) ,  \qquad  \qquad
 \varpi (\phi  , J \phi ) > 0           \label{condJ}    \eeq
 whenever $\phi$ is a real solution.
The latter condition (\ref{condJ}) has an obvious connexion with the need
 to make the sesquilinear form positive on some subspace.
 The former ensures that 
$ \Pi ^+$ is a  symmetric operator with respect to the sesquilinear form 
(\ref{sesq}), which in turn entails that the positive and negative-frequency 
subspaces are mutually orthogonal. 

\medskip  
\noi
The result obtained  by Ashtekar and Magnon \cite{ashmag} for minimal 
coupling  consists in  the  construction of   an admissible
 $\disp  J _  {\Sig}$
 for every  Cauchy surface  $\Sig$.
This was  possible  because,  assuming that spacetime is  globally  
hyperbolic, each solution to the KG equation can be uniquely determined by its 
value  and that of  its normal derivative  on  $\Sig$.   In fact, owing to 
the global generalization \cite{ellhawk} of   Dronne's theorem,  this 
property of the KG equation holds true for a wide  class of hyperbolic second
 order partial differential equations; in particular, it is true also for the
 KG equation with a source term (see Appendix 1).

Given one isometric transformation  $T$   of spacetime,  we are led to 
investigate under which conditions  $J _  {\Sig}$ 
 is actually invariant by $T$.  Naturally, we find that this is the case 
when   $T$  maps    the Cauchy surface $\Sig$ into  itself (Appendix).
For some particular spacetimes, one can find a Cauchy surface invariant by
{\em  the whole group}  of isometries. 
 But  we cannot always expect this situation.
For instance, this is impossible as soon as $(V , g )$  admits a 
timelike Killing vector.      In that case  however,  it is possible  to define 
directly a splitting of the wave functions according to the sign of the 
energy; the projection operator associated with this spplitting obviously 
corresponds to an appropriate  complex-structure operator $J$.

\subsection{The positive-frequency kernel}

\noi Insofar as spacetime is globally hyperbolic, the
  retarded and advanced Green functions and (by difference) the Jordan-Pauli 
commutator "function"  are unambiguously determined by the geometry.

\noi          More problematic is  the kernel  $D^\pm$
 which  (if it exists)  defines  the   
 positive-frequency (resp. negative-frequency) solutions to  the KG equation, 
through the formula
\beq  (\Psi ^\pm) (y) =  \bigl( D^\pm (y,x)  ; \Psi (x) \bigr) = 
          \int j^\alp \bigl(  D^\pm (y ,x)  \Psi (x) \bigr)   \  
d\Sigma  _\alp      
\label{defD+}    \eeq
where  $\Psi ^\pm  =  \Pi ^\pm    \  \Psi $
 is the positive-frequency part of $\Psi$,
 and $y$ is an arbitrary point of $V$.   We adopt a generalized Eistein 
notation: in expressions of the form  $ (.;.)$  we make the convention that 
integration must be performed over the variable which is twice repeated.

\noi       It is clear that $D^\pm$ is a kernel for  $\Pi ^\pm$.
 Note  that $D^\pm$ is a "two-point function" (actually, a 
distribution) and must satisfy the KG equation in both arguments.
Our interest for this kernel is motivated by  the fact that 
 the field operator must be defined through   creation and anihilation 
operators   associated  with  the one-particle state  $D^+$.
 
\noi Naturally, when $D^+ $ exists, we can always recover the projector 
$\Pi ^+$, for  $\Psi ^+ $ is given by (\ref{defD+}) . But the converse 
requires some care: given a complex-structure operator, or equivalently the 
projector  $\Pi ^+$,  it is tempting  to  formally write   (\ref{defD+})
 and claim that it defines $D^+$ as a distribution; indeed associating 
$ \Psi ^+$  to $\Psi$  obviously defines a linear functional.
But in order to make up a distribution, this functional should additionally be 
in some sense {\em continous},
 which requires  some topology on $\calk$.
However, as much as possible, we postpone topological considerations 
  to the end of this study and simply  use 
  the  theory of linear operators  in a vector space.

\noi Most results available in the litterature  are formulated purely in 
terms of  $J$ or $\Pi ^+$, and disregard the kernel. 
In contradistinction, C. Moreno has  exhibited  complex structures  by  
directly constructing  their  kernel.

\section{Warped spacetimes}

\noi  A warped spacetime is a product manifold  $V = V_1 \times V_2$, endowed 
with a metric (omitting indices)
\beq  g = \alp \oplus (-S) \gam       \label{defwarp}        \eeq
 where  $\alp$ and $\gam$, 
respectively, are metric tensors on $V_1 $ and $V_2$, and $S$ is  a positive 
function on $V_1$. Usually $\sqrt{S} $ is referred to as the warping factor, 
and one  sets  $ S = \exp (2 \Theta )$.
                                                            
\medskip     
\noi
 In this work we consider only warped spacetimes of Type I, 
in other words  $\alp $ is Lorentzian and $\gam$ is Riemannian.

\noi  Respective dimensions of $V_1, V_2$  are $p$ and $q$.
 Geometric elements associated with $V_1 , V_2$ are respectively 
labelled with indices $1, 2$. For instance $\Del _2$ is the Laplace-Beltrami 
operator corresponding to $ (V_2 , \gam )$,  etc.
Indices $A, B$  label coordinates in  $V_1$ (resp. $i, j$  in $ V_2 $ ). 
For  $x$  and $y$ in  $V_1 \times V_2$, instead of the canonical co-ordinates
$x^A , x^i$ we can  use the intrinsic notation

$ x =  (u, \xi )  , \qquad   y = (v, \eta ) \quad  $       with 
$ u, v  \in  V_1 ,  \qquad   \xi , \eta  \in  V_ 2 $.

\medskip \noi
The warping is just a geometric structure; it arises in a lot of  different 
physical situations.  For instance the FRW universe  of cosmology (and its 
anisotropic generalizations), any  metric with  spherical symmetry
(stationary or not), and also the bulk spacetimes of brane theory,  are warped. 
The most  simple nontrivial  and nondegenerate  example of warping is given
by the Friedman-Robertson-Walker line element.

\noi  Let us summarize the nice properties of warped spacetime:

a) to a large extent, we  control their isometries.

b) the motion of a test particle in any warped spacetime admits a first 
integral  of particular interest; this remark  can be made already at the 
classical level. In its quantum mechanical version, it provides us with an 
observable which can be diagonalized together with the Klein-Gordon operator.
This situation allows for a mode decomposition of the solutions to the KG 
equation. 

c) Each mode  is a tensor product, which results in a separation of the 
variables;
the skew-symmetric form defined on a given mode admits a  factorization. 

\medskip
\noi
Unless otherwize specified, throughout this paper we make this technical 
assumption:

 {\sl We assume that $V_2$ is compact and connected.}

Remark. We are mainly interested in cases where spacetime is warped in a 
unique way, up to trivial re-definition.
Definition. A trivial re-definition of the warped structure  (\ref{defwarp})
consists in the replacement of $S$ and $\gam$ respectively by
$$ S' = \rho S , \qquad \quad  \gam ' = \rho ^{-1} \gam $$
which results in the replacement of $\Del_2$ by
$\Del ' _2 = \rho \Del _2$ whereas all the eigenvalues are multiplied by 
$\rho$, say  $\lam ' _n =  \rho \lam _n$.  The equation (\ref{reduphi})
remains unchanged, for  $\lam ' _n { S '} ^{-1} =    \lam _n  S^{-1} $.
Moreover $\sron_n ,  \eron _n$ and the operator $D$ remain unchanged.

\subsection{Isometries of warped spacetimes} 

In this section we summarize relevant results of J. Carot and J. da Costa 
\cite{carot}                     and M. S\'anchez \cite{sanch}
                
\noi
The product structure of  $V_1 \times V_2 $ allows for the canonical 
decomposition of any vector field $X$  into its first and second pieces,
 say  $X = X_{(1)}  +  X _{(2)}$,
with    $X_{(1)}  =  X^A  \dron _A ,   \qquad   X_{(2)} =  X^j   \dron _j $.

Remark:  This decomposition does not mean that $X _{(1)}$  (resp.  $ X_{2} $)
might  always be regarded as a vector field on $V_1$ (resp. $V_2 $).
 But  this situation  happens in some  particular cases of interest, see 
below.  Besides, if we fix $\xi$ (resp. $u$) we can regard $X_{(1)}$ (resp. 
$X_{(2)}$) as a vector field on   $ V_1 \times \{ \xi \}$
 (resp.   $  \{ u \}  \times V_2 $ ).

\noi    We can  distinguish  three cases:

i)  First pure  case:  $X  = X_{(1)} , \qquad  X_{(2)}  = 0$

ii)  Second pure case:   $X =  X_{(2)} , \qquad    X _{(1)} = 0$

iii) Mixed case:    $X$  has both pieces nonvanishing.

\noi
The following result has been proved in references \cite{carot} \cite{sanch}.

\noi 
{\bf Theorem 0} ({\sl Carot and da Costa 1993, M. S\'anchez 1998}).

    {\sl In the first pure   case, $X$ is Killing for $(V, g)$   iff
$X_{(1)}$   is Killing for   $ (V_1 , \alp ) $ and in addition we have
$X^A \dron _A   S  = 0 $

In the second pure  case, $X$ is Killing for  $(V, g)$ iff
   $X_ {(2)} $   is  Killing   for  $ (V_2 , \gam ) $

\medskip

In general,    If $X$ is Killing for   $(V, g)$, then 
$X_{(1)} $  is   Killing  for  $ (V_1 , \alp ) \times  \{ \xi \}  , 
 \quad     \forall    \xi  \in  V_2 $,   and
   $X_{(2)}$ is a Conformally  Killing vector for
  $  \{ u \} \times  ( V_2 ,  \gam ) , \quad  \forall      u  \in  V_1 $}.
 (the converse may not be true!)  

\medskip
\noi
In fact these results can be directly derived from the formula  
(\ref{defwarp}) if we consider  $\alp$ and $\gam$  as particular tensor 
fields on  $ (V, g)$. 
Hint:    split the Lie derivative
 operator as ${\rm L}_X = {\rm L}_1 + {\rm L}_2$   where 
$1, 2 $ refer to the vector fields    $X_ {(1)} ,  X_ {(2)}  $ respectively.
Observe that the well-known  general formula
\beq   {\rm L}_Z  t_{\mu \nu} = Z^\sig \dron  _\sig    \   t_{\mu \nu}
+  t_{\sig \nu} \    \dron _\mu    Z ^\sig   
+       t_{\mu \sig}      \      \dron_ \nu   Z  ^\sig     \label{Lie}     \eeq
entails
 \beq     {\rm L} _ 1  \gam  =   {\rm L} _2  \alp   =  0                   \eeq
We obtain    that $X$ is Killing for $(V, g)$ when
\beq   {\rm L}_ 1  \alp  -  ({\rm L}_1 S) \gam  -  S \    {\rm L}_2   \gam 
   = 0                                    \label{kill}         \eeq
Note  that  $ ({\rm L}_ 1  \alp )_{ij}$  and  $({\rm L}_2  \gam )_{AB}$
 always vanish.
 In  contradistinction  $ ({\rm L}_1 \alp )_{Ai}$  and 
    $ ({\rm L}_2 \gam )_{Ai}$    may be  different from zero.
Actually, it stems from (\ref{Lie}) that
 $$    ( {\rm L}_1  \alp  ) _ {Ai} =    \alp _{A \sig}  \dron _i  X_{(1)} ^\sig
   + \alp_{\sig i}    \dron _A  X_{(1)}  ^\sig                        $$   
  $$    ( {\rm L}_1  \alp  ) _ {Ai} =    \alp _{AB}  \dron _i  X_{(1)} ^B
   + \alp _{B i}    \dron _A  X_{(1)}  ^B                              $$
But            $  \alp_{B i}  $  is zero, thus
\beq     ( {\rm L}_1  \alp  ) _ {Ai} =  
         \alp _{AB}  \      \dron _i  X_{(1)} ^B     \label{Ai}        \eeq
and similarly
\beq     ( {\rm L}_2  \gam  ) _ {Ai} =  
         \gam _{ji}  \      \dron _A  X_{(2)} ^j     \label{Aibis}       \eeq
 So 
\beq      ( {\rm L}_1  \alp )_{ij}  =  0     \label{l1alp}   \eeq
\beq ( {\rm L}_2  \gam )_{ij}  = 
 -  ({\rm L}_1  \log S) \    \gam _{ij}  \label{expans}         \eeq

\medskip  \noi
  In the first pure case, equation  (\ref{kill}) reduces to 
${\rm L}_1 \alp  = 
 ({\rm L} _1 S)    \gam  $. Taking the $ij$ components of this formula 
yields ${\rm L} _1 S = 0$, which entails $ {\rm L}_1 \alp = 0$.
 Thus in particular
  $ ({\rm L} _1 \alp) _{Ai} = 0 $.  As $\alp _{AB}$  is invertible, 
(\ref{Ai}) tells that $X^B$ depends only on the variables $x^A$. 
So $X$ is a vector field on $(V_1 , \alp )$.

\medskip \noi
In the second pure case, it is clear that  ${\rm L}_2 \gam = 0$.
 We can write in particular              $ ({\rm L}_2 \gam) _{Ai} = 0 $.
 As $\gam _{ij}$ is invertible, (\ref{Ai}) tells that
  $X^j$ depends only on the variables $x^i$. So $X$ is a vector field on 
$(V_2 , \gam )$.

\medskip     \noi
In the mixed case, $X_{(1)}$ (resp. $X_{(2)}$)
 may also depend on  $x^i$ (resp.$x^A$).
Still, taking the $AB$ components of (\ref{kill}) yields 
 $({\rm L}_1 \alp ) _{AB}=0$, in other words
$$  X_{(1)}^C  \dron _C   \alp_{AB}  + \alp_{AC}  \dron _B   X_{(1)} ^C 
                +  \alp_{CB}  \dron _A   X_{(1)} ^C      =0              $$
so, for fixed $\xi$ with coordinates $x^i$ in $V_2$,  $X_{(1)} ^C$
 is Killing of 
$(V_1 , \alp )$. Taking the  $ij$  components  of (\ref{kill}) yields
\beq ({\rm L} _2 \gam ) _{ij} =
 -  ({\rm L} _1 \log S) \    \gam_ {ij}         \label{severe}      \eeq
So, if we ignore its dependence on the  coordinates $x^A$, we can assert that 
$X_{(2)}$ is conformally Killing for the metric $\gam _{ij}$.
Multiplication of the above formula by $\gam ^{ij}$ and contraction of the 
indices provide 
\beq   \gam ^{ij}  ({\rm L}_2 \gam ) _{ij} = -  q  {\rm L}_1 \log S    
                                    \label{mezig}    \eeq
Remark: This formula is (trivially)  satisfied also in the first pure case.

\medskip
\noi
 In view of these results, we could say, loosely speaking, that in practice
"most"   Killing   vectors are inherited from the symmetries of  the second
 factor manifold, whereas the occurence of the first pure and mixed cases is
  somehow  "exceptional".
This point can be made more precise, as follows.

\noi
In the second pure case, we observe that $X=X_{(2)}$ is Killing for
 $(V_2 , \gam )$  irrespective of the warping factor. In this case, keeping 
$X, \alp, \gam $ fixed, we can arbitrarily replace $S$ by another positive 
function $S'$. It is obvious that $X$ remains Killing for the metric
$ g' = \alp \oplus (-S') \gam$.

\noi In contrast,  in the 1st pure case, $X=X_{(1)}$
 may remain Killing for $g'$
 only if ${\rm L}_1 S'$ vanishes, and in the mixed case, $X$ may remain
 Killing for $g'$ only if $S'$ satisfies the necessary condition (\ref{mezig}).

\noi 
                 
\medskip

 \noi     {\bf Definitions}.  

A Killing vector of $ (V, g ) $ will be called {\sl structural} when it is the 
lift of a Killing vector of $ (V_2 , \gam )$, say $X=X_{(2)}$.

\noi Remark.
            Structural isometries are preserved, if we arbitrarily 
modify the warping factor, leaving the factor-metrics
 $\alp$ and $\gam$ unchanged.

A Killing vector $X$ of $ (V, g )$  will be called {\sl factorial} when 
$X_{(1)}$ is either zero or Killing for $(V_1 ,\alp )$  whereas 
$X_{(2)}$ is either  zero or           Killing for $(V_2 ,\gam )$

A Killing vector $X$ of $ (V, g )$  will be called {\sl  extraordinary}  when 
$X_{(2)}$ is a  {\em properly} conformal  vector of
  $ {u} \times  ( V_2 , \gam )$  for all $u \in V_1$ (by properly conformal 
we mean that it  is not an isometry).

The following statements can be read off from Theorem 0.

{\bf Corollary} 

{\sl  When  $(V_1 , \alp ) $  has no Killing vector,
 then every Killing vector of  $(V_, g)$   is structural.

When       $(V_1 , \alp ) $ admits  infinitesimal isometries,
the existence of a non-structural symmetry  in $(V, g )$ still requires
 a particular shape of the warping factor, expressed 
by  formula      (\ref{mezig}).}

\medskip
The  definitions given above help to classify the isometries as:
structural, factorial, extraordinary.
In this paper we limit our investigation to warped spacetimes that are free 
of extraordinary Killing vector.

\noi   {\em   Under this restriction}, in view of Theorem 0,
 we can assert that if $X$ is Killing for 
$V$,   we simply have $X=X_{(1)} + X_{(2)}$ where, on the one hand
     $X_{(1)}  $  is  Killing for $ (V_1 , \alp )  $       and in addition
$ X_{(1)} ^A  \dron _A   S  = 0$.        On the other hand
$X_{(2)} $ is  Killing for   $ (V_2 , \gam )   $.

In general it remains possible to have a Killing vector $X$ such that
$ X_{(1)} ^A  \dron _A   S $  vanishes.  We shall only consider the  cases 
where $X$ is 

either $ \     $      a) always timelike               $ \qquad   $     
    or   $\        $            b)  always spacelike.

$$ \                 $$

\noi  {\bf Special cases}

  $p=1$

\noi This case  describes  a generalized FRW  spacetime,    one has  
 $ V_1  \subset  \batonR$. Existence of  a non-structural Killing vector for
  $ (V, g )  $,  requires an exceptional shape of the warping factor.    

\noi
 The exceptional warping factors are determined by an ordinary differential 
equation $ \disp     \ddot {\Theta } \exp (2 \Theta) = {\rm const.} \     $ 
in terms of a suitable time scale $t$.
 They are listed  in the literature  and include, of 
course, the de Sitter universe. In fact the de Sitter metric
 is {\em a priori} exceptional  because its full isometry group is larger than
 the isometry group  of one of its space sections.

\medskip
\noi    $p=2$

\noi    Assuming that the lines defined in $V_1$ by 
  $S= {\rm const.}$  are everywhere   timelike (resp. spacelike). 
This assumption excludes the trivial case of a constant  warping factor.

\noi  The shape of the warping factor induces on $(V_1 , \alp )$  a preferred 
net of orthogonal coordinate lines, and a preferred foliation by Cauchy 
 "surfaces"  wich  are in fact  the lines  $S= {\rm const.}$ (resp. their 
orthogonal trajectories).

\medskip
\noi {\sl Four-dimensional spacetime $p=q=2$}

\noi This situation  is referred to as Class B in ref. \cite{carot}. 
It  encompasses all spacetimes with spherical symmetry.

\section{Mode Decomposition}
\subsection{An interesting constant of the motion}

At the classical level, it is remarkable that the phase-space function 
$2K_{\rm cl} =  \gam ^{AB}  p_A p_B $ is invariant by action of the
 geodesic flow. The quantum mechanical version of this result states  that
the differential operator $2K_{\rm qu} = - \Delta _2$  commutes \cite{droz} 
with the KG 
operator for minimal coupling, namely   $  \nab ^2 + m^2$.
Indeed, the structure of warped spacetime allows to derive  a useful identity,
\beq    \nab ^2   \   \Psi  =  \Delta _1 ^\sharp \    \Psi   -
S^{-1} \      \Delta_2   \   \Psi      \label{reduc}             \eeq
where    $ \Delta _2  $  is the  $q$-dimensional Laplacian 
 associated with the manifold  $ (V_2 ,  \gam )$   and where we define in 
$(V_1 , \alp )$ the "warped Laplacian"  of a function
$$ \Del _1 ^\sharp   \Psi =
{1 \over   \sqrt {|\alp |}  }  S ^{- q / 2}
\partial _A (    \sqrt {|\alp |} \     S ^{q / 2}  \ 
  g^\AB  \partial _B  \Psi )                                              $$
which can be re-arranged as
\beq  \Delta _1 ^\sharp \     \Psi =  \Del _1  \Psi +
\half  q  \alp ^{AB} ( \dron _A \log S) \dron _B \Psi   \label{defD}    \eeq
irrespective  of whether  $\Psi$ is a solution to (\ref{weq})  or not. 

\noi
Note that $ g^\AB = \alp ^\AB$, 
thus  the second order differential operator  $\Delta _1 ^\sharp$
  only affects quantities  depending on the $x^A$ variables. 
In contrast,   $ \Delta _2$, as an operator extended to functions  on $V$,
 does not affect the quantities depending on $x^A$ only.

\medskip
\noi  {\bf  Examples }

\noi  When $(V, g)$ is FRW with flat space sections,  $ (V_2, \gam ) $ is the 
three-dimensional plane, and  $K /m$  is the kinetic energy. In this 
particular case, conservation of $K$ is trivial: space translation invariance  
implies that each $p_k$ is conserved. For FRW with spherical space sections, 
 $ (V_2, \gam ) $  is a three-dimensional sphere, in principle, conservation 
of $K$ could be also derived from the constants of the motion associated
 with the Killing vectors on the sphere. When $(V, g)$ is some inhomogeneous 
and anisotropic generalization of FRW,   $ (V_2, \gam ) $ has perhaps no 
isometry  at all, but  $K$ is still conserved. In that case, it is  natural
 to consider   $K/m$ as a generalization of the  kinetic energy.

\noi When $(V, g)$ is four-dimensional with spherical symmetry,
 $(V_2 , \gam )$ is the sphere with unit radius and line element
$$   \gam _{ij} dx^i  dx^j      =
d \theta ^2  + \sin ^2 \theta \    d\varphi ^2              $$ 
 We find, with the standard notation, that 
$$ 2 K_ {\rm qu} =  \Del _2 = 
{1 \over \sin \theta }   {\partial \over \partial \theta}
  (  \sin \theta \    \       {\partial \over \partial \theta})  +
 ( {1 \over \sin \theta }  \        {\partial \over \partial \varphi} )^2  $$
which is opposite to the square  of angular momentum.

\medskip
\noi These examples indicate that, in most cases of interest, $K$ has a
natural physical  meaning.

\subsection{Mode Solutions, Separation of Variables}

For all $\lam \in {\rm Spec}  V_2$ the {\em Mode Space} associated 
with $\lam$ is 
by definition the linear space $\hron [\lam ] $  made of solutions to 
(\ref{weq}) that are also 
eigenfunctions of  $-\Delta _2$  with eigenvalue  $\lambda $.  
Since we assume that $V_2$ is compact and connected, $ {\rm Spec} (V_2)$
is a discrete sequence cite{geomspec}
$$ 0 < \lam _1  \cdots <  \lam _n  \cdots   $$ 
We simply write   $\hron _n$  as 
a short-hand  for       $\hron [\lam _n ]$.
This terminology agrees with the usual one in the special case of FRW 
spacetimes.

    \noi
Let us now turn to features associated with  $(V_2 , \gam )$.
The coordinates $x^A$ are ignorable when  $\Del _2$ acts  on $\Phi $, so let  
 $\eron [\lam ] $ be the (complex) eigenspace of  $- \Del _2$ in 
  $C^\infty (V_2 , \gam )  $. We write $\eron _n$ as a short-hand for 
$\eron [\lam _n ]$

\noi Note that $\eron$ has finite dimension $r (n)$ and admits a real basis
\cite{geomspec}.

\noi   Since  $(V_2 , \gam )$ is elliptic,  $C ^\infty (V_2 )$
        (thus also $\eron _n$)
 is endowed with the  {\em positive definite}  scalar product
\beq <F , H >_2 = \int _ {V_2} F^* H  \  \sqrt{ \gam} \       d_2 ^q x
  \label {scalpositif}                                                   \eeq   
For  $F$ and $H$ real,  $ <F, H>_2 $ is a positive quadratic form.

\noi  Let  $\disp      E _{1,n} , \cdots    E _ {a,n}   \cdots E _ {r,n} $
  be a real orthonormal basis  of $\eron _n$, 
thus  $<E_a , E_b > _2 = \del _{ab}$.

\medskip

\noi       We focus of these solutions to (\ref{weq}) that are a
  {\em finite} sum of modes, say
\beq        \phi = \sum \Phi _n        \label{summodes}         \eeq
The loss of generality resulting from this restriction to finite sums will,
 to a large extent, be  compensated when ultimately  performing the
 Hilbert completion of the positive-frequency sector.

It is important to check if  the modes that we have defined actually are 
orthogonal. So, the question arises as to know whether  
 $\hron_n$ is orthogonal to $\hron_l$ for $l \not= n$,
 in the sense of the sesquilinear form  associated with the Gordon 
current~\cite{droz}.
In fact the answer is: yes  (see \cite{droz} Section 5, Proposition 5).

\medskip
\noi  
As  long as    $\lam _n$  is kept fixed,  the label
 $n$ referring to a determined eigenvalue of  $K$ can be provisionally dropped.
Thus $\Phi$ stands for   $\Phi_n $   in Mode $n$. 
For some nonnegative $\lambda  \in {\rm Spec}  (V_2) $ we have
\beq \Delta_2  \Phi  =  - \lambda   \    \Phi     \label{eq:lapl}  \eeq
and (\ref{weq}) reduces to
\beq (\Delta _1 ^\sharp + \lam   S^{-1} + m^2 )\     \Phi   = 0 
                                \label{reduphi}      \eeq
The coordinates $x^k$ are ignorable when  $\Delta _1 ^\sharp$
 acts  on $\Phi $, so the 
equation above is a  $p$-dimensional problem.
Let  $\sron $ (resp. $\sron ^R$) be the linear space of $C^\infty$ 
complex-valued (resp. real-valued) functions
 $f (x^A)$ satisfying the equation
\beq (\Delta _1 ^\sharp + \lam   S^{-1} + m^2 )\     f  = 0  
  \label{reduf}          \eeq
We have shown \cite{droz} that     (\ref{reduf}) can be cast into the form
\beq  (\Del _1 + \Xi )\    \fhat = 0        \label{redufbis}      \eeq
where we have set
\beq   \disp     \fhat = S ^{q/4} \    f         \label{rescalf}  \eeq
 \beq     \Xi =      (S ^{q / 4}  \  \Del _ 1  S ^{-q/4} )    -
{q^2  \over 8}\   S^{-2}  \    \alp (\partial S , \partial S )
+  \lam  S ^{-1} +   m^2                           \label{explixi}    \eeq
On the one hand  (\ref{redufbis}) is simpler than the original 
equation (\ref{weq}) because it is only a $p$-dimensional problem; on the 
other hand it seems to be  a little  more complicated, for it involves
 a "source term"$\Xi$. 

\noi    Fortunately, the KG equation with a source term 
still admits a conserved current\footnote
{Notice a slight change of notation with respect to ref. \cite{droz},
 because here $J$ denotes the complex structure.},
 which entails that  the quantity
\beq     \sig (\fhat , \hhat ) =
\int _L                             ( \fhat  \alp ^{AB} \partial _B  \hhat 
- \hhat   \alp ^{AB} \partial _B   \fhat  )     \quad  dL_A
                                    \label{skewredu}        \eeq 
if it is finite, doesnot depend on the spacelike $ p-1$ dimensional 
surface  $L$.
Let us ensure that $\sig$  actually is finite; when  $L$ is not compact
 it becomes necessary that  the functions $f$ and $h$
 decrease rapidly enough at spatial infinity.
In order to be more precise,  we make either one of the  following assumptions:

\noi {\bf Assumption 1} 
    
{  \sl For  $p >1$, we suppose 
$ (V_1 , \alp )    \approx  \batonR  \times {\rm compact}$.   }

\noi {\bf Assumption 2} 
    
{\sl For  $p >1$, we suppose that 
   $ (V_1 , \alp ) $ is globally hyperbolic, which implies some  Cauchy 
surface $ L \in V_1$. Moreover we  include in the 
definition of $\sron$  the condition that $f$ in equation 
    (\ref{reduf})  arises from initial data with compact support.}\footnote
{This property, when it holds, is independent of the choice of the Cauchy 
surface (see Ref. \cite{wal} pp. 53-58).   } 

Note that none of these assumptions contains the other one,
 although they overlap. From now on,  we suppose that  one of them is realized.
Only if necessary, shall we specify which one.

Remark: 

 \noi   Strictly speaking, spherically symmetric models that arise in 
astrophysics fail to satisfy these assumptions, for they have 
$ V_1    \approx   \batonR  \times \batonR ^+ $.
But it is possible to enlarge the present framework with help of a 
prescription: extending the Cauchy "surface" $t=0$ to negative values of $r$
and making the convention that $f (t, -r)    =  f(t, r)$.

\medskip  \noi
Now we can endow  $\sron ^R _n$  with the skew-symmetric form
\beq     W (f, h ) =     \sig (\fhat , \hhat )     \label{defW}           \eeq
With help of    (\ref{rescalf}) we can check that
\beq W (f,h) =  \int _L  S ^{q/2}
 ( f  \alp ^{AB} \partial _B h 
- h  \alp ^{AB} \partial _B   f  )    d L_A   \label{sskew}           \eeq
as we had written in ref. \cite{droz}. Extending $W$ to the complex domain,
we define a sesquilinear form
\beq    (f ; h ) _1        =   -i  W (f ^*  , h )              \eeq

{\sl Remark}

\noi When $p=1$ then $W$ reduces to the Wronskian of two functions 
of a single variable $x^0$.
 This situation occurs in generalized FRW spacetimes.

\medskip

\noi     The general solution to   (\ref{eq:lapl})(\ref{reduphi})
 takes on the form
\beq  \Phi  =    \sum _1 ^{r(n)}    f _a (x^A) \    E _a (x^j)  
                \label{devphi}        \eeq
where   every  $f _ a$  is a smooth function  on    $V_1$, satisfying 
equation   (\ref{reduf}), and the mode label $n$ has been dropped from 
$E_{a,n}$.   In other words we can write
\beq    \hron _n   =   \sron _n   \otimes \eron _n   \label{factor}  \eeq

\medskip

 It is clear  that the sesquilinear form $ (\Phi ; \Omega )$  and the 
 skew-symmetric form  $\varpi$  can be restricted  to $\hron _n$.

\noi {\bf Definition}

\noi Let $w$ be the {\em restriction}  of $\varpi$ to $\hron _n$.

\medskip \noi
{\bf Theorem 1} {\sl  If $(V_1 , \alp )  $ is globally hyperbolic, then all the 
solutions of the form   (\ref{summodes}) are uniquely determined by their 
Cauchy data on $L \times V_2$ where $L$
 is a Cauchy surface in $( V_1 , \alp )$}.

The proof is mode-wise. For $\Phi $ in mode $n$, we write   (\ref{devphi}) and 
observe that 
$$ \Phi _n  |_\Sig  =   \sum _a  f_a (u) |_L  \    E_a (\xi )    $$
$$   ( \dron _0 \Phi _n ) | _ \Sig  =
    \sum _a   ( \dron  f_a ) |_L  \    E_a (\xi )               $$          
But $\fhat _a$ is solution to a reduced
 KG equation 
  (\ref{redufbis}); it follows that  $\fhat$ and therefore $f$, is uniquely 
determined by its Cauchy data on $L$.

At this stage one may be tempted to claim that   $L \times V_2$
 is a Cauchy surface for $ (V , g )$. But we leave this question open.

\noi For  complex functions in  $\hron _n$ we can write in general
\beq     (\Phi ; \Psi ) = - i  w (\Phi ^* , \Psi ) ,  \qquad  \quad
   (f ; h ) _1   = - i W (f ^*  , h)      \label{underskew}   \eeq

\subsection{Factorization}

\noi    
 Let  $\Phi $    and $\Omega $ be  solutions in mode $n$.
So $\Phi $       can be written as       (\ref{devphi})       and 
\beq   \Omega = \sum _1 ^{r(n)}  h _b  E _b        \label{devomeg}       \eeq
Under reasonable assumptions \cite{droz},
 the sesquilinear form $w$ on    $\hron _n$
is {\em compatible}  with the structure of tensorial product.
Indeed for 
$\Phi = f F \in \hron _n$ and  $\Omega = h H  \in  \hron _n$
  with $f, h \in \sron _n$   and  $F, H  \in  \eron _n$,    we have proved 
that
\beq  ( \Phi ; \Omega ) =  (f;h)_1 \   <F,H>_2  \label{factor2}   \eeq
where    $  <F,H>_2 $      is invariant under  the symmetries
 of $ (V_2 , \gam )$.

\medskip
\noi
In order to ensure that  $  (f;h)_1 $ is finite, 
in ref. \cite{droz} we used Assumption I. 
But the reader will easily check that the derivation of (\ref{factor2}) and the 
conclusions of Section 5 in ref. \cite{droz} remain valid as well under 
Assumption II.

\medskip  \noi
Equation (\ref{factor2}) 
can be reformulated in terms of the underlaying skew-symmetric
 forms.
 For $\Phi$ and $\Omega$ as above, then (\ref{underskew})  entails that  
$$ (\Phi ; \Omega ) =  - i w (\Phi ^* ; \Omega ) $$
Our result of   \cite{droz} tells that 
\beq  w (\Phi , \Omega ) = 
   W (f , h ) \       <F , H >_2            \label{factoriz}    \eeq
Here we  specify  $F =  E_a, \quad   H =  E_b$ and  obtain that
$$ (\Phi ; \Omega ) =  \sum   (f_a ; h_b )_1 \      \del _{ab}       $$
In view of  equations   (\ref{underskew})    we can re-write 
this in terms of skew-symmetric forms,  say
  \beq   w  (\Phi ^* , \Omega ) = 
      \sum _ {ab}  \      W (f_a ^* , h _b ) <E_a , E _b > _2      \eeq
 (For each mode $\hron _n$)  it is now clear that   $w$ is the 
skew-symmetric form      induced  by  $W$  on  $\hron _n$ according to the 
tensorial  product structure of  it.

\subsection{Modes, Complex Structure  and  Invariant Splitting}

We turn back to the problem of determining a  complex structure $J$ in the 
full space of solutions to (\ref{weq}).
If we succeed, the positive (resp. negative)-frequency part of $\calk$, say 
 $\calk ^+$ (resp. $\calk ^-$)  is
made of solutions satisfying $ \Pi ^+ \phi = \phi$
(resp. $ \Pi ^- \phi = \phi$).

Insofar as the quantity $K$ distinguished in Section 4.1 has a physical meaning,
 it should be represented by an operator     
which maps $\calk ^+$  into itself.  
This condition can be formulated as follows:

\noi   For all $\phi \in \calk$,  we have $\Pi ^+  \phi \in  \calk ^+$.
Now, $K  ( \Pi ^+ \phi ) $   must enjoy the same property, thus
$ \Pi ^+  K  \Pi ^+  =  K \Pi ^+$.  But $\Pi ^+$ is  idempotent, so we can 
write 
$ \Pi ^+  K   \Pi ^+  =         ( K \Pi ^+  )  \Pi ^+    $, in other words
\beq       [\Pi ^+ , K  \Pi ^+ ] = 0   \label{KPi+}   \eeq
In order to preserve the symmetry between positive and negative frequencies, 
it is natural to require   as well
\beq             [\Pi ^- , K  \Pi ^- ] = 0          \label{KPi-}          \eeq
But $ \Pi ^-  =  1 - \Pi ^+$. Inserting this relation into
  (\ref{KPi+})   (\ref{KPi-})     finally yields
$ [K, \Pi ^+] = 0$. 

\noi  It is now clear that $K$ leaves stable both $\calk^\pm$  iff
$K$ commutes with $\Pi ^+$ or equivalently with $J$.
In order to implement this condition,  the separation of frequencies will be 
mode-wize carred out. 
We are led to focus on the solutions to (\ref{weq}) that can be developed 
in modes,  like $\phi$ in  (\ref{summodes}).
 Let $\cal L $ be the linear space formed by these 
solutions.

\noi It is a mere exercise to check that,  provided $J_n$ is a complex 
strucure in $\hron ^R$,  and $J_n$ is positive with respect to $w$, then the 
operator $ J$  defined by direct sum (say
  $ \disp  J = \bigoplus   J_n$       that is
        $ J \phi  = \sum  J_n \Phi _n$)
 is a complex structure in ${\cal L}  ^R $ and is positive 
with respect to $\varpi$.

\noi
 Now, in each shell $\hron_n$, we look for suitable  projectors
   $\Pi^{\pm} _n$, 
 or equivalently  we look for  a  complex structure $J_n$ 
acting  in    $\sron _n  \otimes \eron _n  $  as a linear operator; let us 
formulate:

\noi   {\bf Problem at a fixed Mode} $n$

\noi  {\sl Find a complex structure operator $J_n$ acting in $\hron _n $, 
positive with respect to $w$  and invariant under the structural isometries.}

\noi    For product solutions  structural isometries act as follows
$$   T ( f (x^A )  F (x^j ) )  =      f (x^A )  T  F (x^j ) $$
where $T$ is an isometric transformation  of  $(V_2 , \gam )$.

\noi   In order to solve the above problem,  we shall resort to  a   few
  results concerning  complex structures in  tensor-product spaces.
They are displayed in the next Section.

 \section{Skew-symmetric forms and complex structure in a tensor product}

{\bf Definition}  { \em A complex structure} on a {\em real}  linear space
 $\aron$ 
 is a linear operator $J$  such that $J^2 = -1$.  

\noi   In this section,  $\Phi , \Psi , \Omega$ may belong to  $\aron$ or 
its complexified, $\aron ^C$.          

\noi   When $\aron$ is a endowed with a  skew-symmetric form $w$,
 it is noteworthy that $J$ leaves $w$ invariant iff $w (\Phi , J \Psi ) $
 is symmetric under exchange of $\Phi$ with $\Psi$.

\noi   Now, consider {\em real} vector spaces  $\aron _1 , \aron _2$.
Let  ${\cal A} = \aron_1  \otimes \aron_2$, where  ${\rm dim}   \aron _2  < 
\infty $.    We assume that
$\aron _1$ is endowed with a  {\em skew-symmetric  form} $ W (f,h ) $ whereas 
$\aron _2$ is endowed with a  {\em quadratic   form}   $ Q (F,H) $.

\noi  It follows  from our assumptions that ${\cal A}$ is in turn endowed with
 the unique skew-symmetric  form     $ w ( \Phi ; \Omega ) $
such that, if   $\Phi = f \otimes F  $ and  
$\Omega =  h \otimes H  $      are in  $  {\cal A} $ we have
\beq  w ( \Phi , \Omega ) = W (f,h) \  Q (F,H)      \label{prodsymp}    \eeq 
We can  say, in an obvious sense,  that  this skew form is {\em compatible}
                                    with the tensorial product.

\medskip
\noi
In addition we assume that the quadratic form $Q$ is  {\em positive definite}
and we  adopt this notation 
$$  Q (F, H)    =  <F, H>    $$  
Since   $\aron _2$ has a finite dimension, it  admits an orthonormal   basis,
 say $ \{E_a \}$, with $a=1, \ldots {\rm dim.}  \    \aron _2$.
   It is clear that 
 $$ Q (E_a , E_b ) = <E_a , E_b>       = \delta _{ab}$$
 Under these assumptions we can easily check  that
\beprop
Any  complex structure operator, say  $J_1$,      defined on  $\aron _1$  
induces  a  complex structure   $J = J_1 \otimes  1$   defined  on 
  $ {\aron}$. If $J_1$ leaves  $W$ invariant, then  $J$ 
leaves $w$ invariant.       \enprop
Proof:
As   $J$ is characterized by  $ J (fF)  = (J_1 f )  F$ 
 it is obvious that $ J^2 = -1$.   

\noi  Invariance of  $w$ can be first proved for products. In this case,
       $w (\Phi , \Omega ) $ is given by     (\ref{prodsymp}) and we have 
$J \Phi =  (J_1 f)  F  , \qquad  J \Omega = (J_1 h) H$. Therefore 
\beq   w (J \Phi , J  \Omega )   =  W (J_1 f , J_1 h) \    Q (F, H)    \eeq
Since we assume that $J_1$ leaves $W$ invariant,  we can replace  
  $   W (J_1  f , J_1 h)$ by simply $ W (f,h)$, thus
$ w (J \Phi , J \Omega ) =  w (\Phi , \Omega ) $ when $\Phi , \Omega$ are 
products. 

\noi But in  general we must write
\beq \Phi = \sum _a    f_a  E_a  \qquad \quad
     \Omega = \sum _b   h_b E _b                      \eeq
So   $w (\Phi , \Omega ) = \sum _ {a,b}  w ( f_a E_a , f_b E_b ) $. 
Apply formula   (\ref{prodsymp})   with $  F = E_a ,    H = E_b $. We obtain
\beq  w (\Phi , \Omega ) = \sum _ {a,b}   W (f_a , h_b ) Q ( E_a ,  E_b) \eeq
Since $\{ E_a \} $ is orthonormal, it follows that 
\beq    w (\Phi , \Omega ) = \sum _ a    W (f_a , h_a ) \label{wPhiOm}  \eeq
On the other hand, 
$ \disp   J \Phi = \sum _a   J (f_a  E_a ) = \sum ( J_1 f_a ) E_a  $,
 whereas
$\disp    J \Omega = \sum _b   J (h_b  E_b ) = \sum ( J_1 h_b ) E_b  $. 
Hence
\beq w ( J  \Phi ,  J \Omega ) = \sum W (J_1 f_a , J_1 h_b) Q (E_a , E_b ) 
=  \sum  W  (f_a , h_b )  \delta _{ab}   \eeq
which is nothing but expression (\ref{wPhiOm}).

\noi    {\bf Definition}

  Let  $J$ be a complex structure on a linear space  $\aron$ endowed with the 
skew-symmetric form $w$.
 We say that $J$   is  {\em  positive} with
 respect to  $w$ when   it leaves $w$ invariant    and satisfies
$ w (\Phi , J \Phi) > 0  $ for every $\Phi \not= 0$.
   
\beprop
If $J_1 $ is positive with respect to $ W $,  then  $J$ is 
positive with respect to  $ w $. 
\enprop      

Proof    

\noi
We know from Proposition 1  that  $J$  is a complex structure for $\aron$
 and leaves $w$  invariant.      Now let us evaluate         
                                   $ w (\Phi , J \Phi)   $.
\noi      For any $\Phi$ in $\aron _1 \otimes \aron _2$ we can  write
\beq    \Phi = \sum    f_a     E _a               \qquad
      J \Phi = \sum (J_1 f_a ) \       E_a       \eeq
where        $ <E_a , E_b > = \delta _{ab}  $.
$$ w (\Phi  , J \Phi ) = 
\sum w ( f_a E _a ,  (J_1 f_b ) E_b  )    =
\sum  W (f_a , J_1 f_b )  \      < E_a , E_b >      $$
But $J_1$ is supposed to be positive with respect to  $W$. Thus 
each term 
  $ W  (f_a , J_1 f_a )  > 0 $ unless $f_a $ vanishes.  It follows that 
   $  w (\Phi  , J \Phi ) > 0$ unless  $f_a $ vanishes for all $a$, which would 
occur only when  $\Phi =0$.

Remark   At this stage it is worthwile recalling that 
the eigenspaces of $\Pi ^+$  and  $ \Pi ^-  $  are mutually orthogonal.

\section{Invariant Separation of the Frequencies.}

\subsection{Structural Invariance}

Let us now turn to the problem at mode $n$, formulated in Section 4.
Let $\eron _n ^R$ contain the {\em real} elements of $\eron_n$.
We claim that

\noi                {\bf Theorem 2}
{\sl Provided that  $J_{1,n}$ is  a complex structure  in  $\sron _n ^R$  and
is positive with respect to the skew form  $W(f,h)$, then
a solution to the Problem at Mode $ n$    is given by
\beq  J_n  =  J_{1,n}    \otimes  I_{2,n}       \label{J1n}   \eeq
where         $  I_{2,n}$   is the identity on     $\eron _n ^R$}.

Formula  (\ref{J1n}) will be referred to as a {\em canonical solution} to 
the problem at a fixed mode $n$.
It means that for any basis in  $\eron _n$
$$ J_n  \sum  f_a  E_a  =  \sum  ( (J_{1,n}  f_a )  \      E_a        $$
The proof is in three steps, one must prove that:

  i)  $J$ is a complex structure, 

ii) it is positive with respect to the skew form in $\hron$,

iii)  it is invariant under the continuous isometries of $(V_2 , \gam )$.

\noi  In view of  (\ref{factor}), 
points (i) and (ii) stem from application of Propositions 1, 2  with
$\aron _1  = \sron ^R$ endowed with a skew form $W$ as in    (\ref{sskew})   
 and $\aron _2 = \eron ^R$  endowed with     $ <  ,  >_2 $ 
as in  (\ref{scalpositif}). 

 \noi    Finally, invariance under the isometries of    $(V_2 , \gam )$ 
is obvious,  for these transformations affect neither the functions 
of $x^A$ nor the identity on $\eron$. This achieves the proof.

\medskip

\noi By Theorem 2,   the initial problem, formulated in $\calk$ and involving a 
symmetry requirement, has been reduced to the question as to construct in each 
$\sron _n$ a complex structure which is positive with respect to $W$.
This reduced problem does not involve any  symmetry condition and is posed in 
a lower dimension ($p$ instead of $p+q$).
In the special case where $p=1$, it is easy to find the complex  structure, 
because $\sron$ is two-dimensional.
Otherwize, the issue  seems a bit more difficult,  because $\sron$ has in 
general infinitely many dimensions and is defined through a KG equation
 with a source term, $\Xi$.

\noi Let us now prove  that a suitable $J _{1,n}$ as invoked in the theorem 
above actually exists.  Our basic tool is the observation that, in a globally 
hyperbolic spacetime, the KG equation with a source term has a well-posed 
initial value formulation (see \cite{ellhawk}, see \cite{wal} p. 56).   

\noi   $\sron _n ^R$ is defined through equation (\ref{reduf}) or 
equivalently  (\ref{redufbis}).
In  (\ref{redufbis}) the source term  $\Xi _n$, explicitly given by 
(\ref{explixi}),  is a smooth function on $V_1$.
  Since           $(V_1 , \alp ) $ is globally hyperbolic, we can exhibit an 
operator $J_{1,n}$ enjoying the required properties, with help of a Cauchy 
surface in $(V_1 , \alp )$. We  proceed as follows:
Let $ {\widehat \sron} _n $ be the vector space of real solutions to 
(\ref{redufbis}).
Let   $L$ be a Cauchy surface  $(V_1 , \alp )$. By the procedure indicated in 
Appendix we construct a complex-structure  operator $j_n$ acting in
  $  { {\widehat \sron} _n  } ^R$. This $j_n$ is positive with 
respect to $\sig (\fhat , \hhat )$. In view of (\ref{defW}) it is a mere 
exercise to check that 
$$   J_{1,n}  =  S^{-q/4} \    j_n  \     S^{q/4}                        $$
is a complex-structure operator acting in  $\sron _n ^ R $ and is 
positive with respect to $W (f,h)$.

\medskip
\noi
After having ensured existence of  $J_n$ for all $n$, we are now in  a position 
to   define  the subspaces $\sron ^{\pm} _n$.

\noi    {\sl Notation} $  $  We  define the projectors
  $\Pi ^{\pm} _ {1,n}  =  \half   (1 \pm i J_{1,n} ) $.

\noi    Let  $f \in  \sron ^{\pm} _n $ when  $f \in \sron _n$  and  
 $\Pi _{1,n} ^{\pm}  f = f$.

\noi It is clear that  $    \sron  _n   =
    \sron ^{+} _n     \oplus  \sron ^{-} _n$. Moreover     
          $   \sron ^{+} _n  $  and  $  \sron ^{-} _n  $   are mutually 
orthogonal  in  $ (. ; . )_1 $.

\noi
Similarly we define 
$$ \hron ^{\pm} _n =         \sron ^{\pm} _n  \otimes  \eron _n $$
we obtain 
  $  \hron _n  =  \hron _n ^+   \oplus   \hron _n ^- $.
and we can say $ \Phi \in  \hron _n ^{\pm}$  iff  $\Phi \in  \hron _n$  and  
  $ \Pi ^{\pm}  _n  \Phi   = \Phi  $.

\noi Let us now achieve our goal, considering generic solutions to 
(\ref{weq}) in the form of finite sums like 
$\Phi = \sum \Phi _n  , \    \Phi _n   \in   \hron _n$.
In other words  $\Phi \in \hron$ where 
$ \disp    \hron = \bigoplus _{n=0}  ^\infty    \hron _n     $.
From $J_n$ and  $\Pi ^{\pm} _n$, by direct sum, we construct $J$ and 
$\Pi ^{\pm}$  acting in the full space $\hron$, according to 
$$    J \sum \Phi _n =  \sum  J_n \Phi _n ,   \qquad \qquad
      \Pi ^{\pm}  \sum \Phi _n  =  \sum \Pi ^{\pm}   _n    \Phi _n    $$
We end up with  $\hron = \hron ^+  +  \hron ^-  $ where 
$\hron ^{\pm} = \bigoplus  \hron ^{\pm}  _n $, and  $\hron ^+  $  orthogonal 
to $\hron ^- $.
Here, $\hron$ endowed with $(.;.)$ is only pre-Hilbert.
 Its completion will provide 
a Hilbert space of one-particle states with positive frequency.

\noi We summarize

\beprop   
 Provided that    $ (V_1 , \alp )$ is globally  hyperbolic,
 there exists a complex-structure operator $J$ which is positive with respect 
to the skew form  (\ref{defw}) and  commutes with structural isometries.
\enprop

In other words, a  splitting   of  the  solutions to equation (\ref{weq}) 
 according to positive and negative frequencies   is possible  and 
 invariant under  structural isometries.

\subsection{Non-structural Isometries}
\noi The previous  result is satisfactory insofar as  $(V_1 , \alp )$
 has no Killing vector.  But our purpose was to investigate  a little further.
So, let us consider the  cases referred to as  a)  and  b)  in Section 3.

a)  $\qquad  X=X_{(1)}$ globally timelike,  $(V_1 , \alp )$ is stationary.

\noi  There are  coordinates where $X^A  \dron _A   =   \dron _0$. 
But there is no evidence that the operator $J_{1,n}$ 
  build  as in  the previous subsection  would be invariant under this $X$.
Under Assumption I,
 we propose an alternative choice,  more direct and more  
natural. 
Every $f$ in $\sron _n$ will be developed as
\beq    f= \int _{- \infty}  ^{+ \infty} f_E dE                      \eeq
where 
\beq Xf = i E f_E                   \label{eigenE}  \eeq
We define 
\beq   \Pi_{1,n}  ^+ \       f  =  
              \int _0  ^{+ \infty} f_E dE                      \eeq
and   similarly   $  \Pi_{1,n}    ^- $. It follows that
     $ \Pi_{1,n}  ^+$  commutes with $\dron _0$.
In agreement with  this modification,  we propose the complex 
structure defined by
$ \disp  J _{1,n}  =  -i (2 \Pi  _{1,n}  ^ +     - 1)$.
It is easy to verify that this operator leaves $W$ invariant and is positive 
with respect to $W$.
 Hint:    $[\Pi_{1,n}  ^+ ,  X ] = 0$ and $X \fhat=     S^{q/4}  Xf$.
From the  new    $ J _{1,n}$ the formula of Theorem 2 yields a 
new $J_n$ which remains of course invariant under structural isometries, but 
is additionally invariant under the time translations.

b)   
    $ \qquad  X=X_{(1)}$ globally spacelike. 

\noi All we need is to  manage that 
the operator  $ J _{1,n}$ be invariant under $ X=X_{(1)}$. It would be 
suficient that, in the procedure described in Appendix, we chose $L$ to be 
invariant by action of $X$.  Still the question arises: does it exist in 
$(V_1 , \alp ) $ a Cauchy surface invariant by the isometries of this manifold?
Naturally, this question is more easy to handle  in two dimensions.

Remark:

\noi
For applications, the  case where  $p=2$ is of particular interest.
 Indeed every two-dimensional 
spacetime is locally conformal to Minkowski; thus whenever  
 $ (V_1 , \alp )$) is globally conformal to a two-dimensional Minkowski space,
 it is itself globally hyperbolic \footnote{Global hyperbolocity only 
involves the causal structure; it is preserved by a conformal factor.}.

\section{Kernels}

\noi
Consider a functional space $\fron$  endowed with a sesquilinear form
noted   $ (.;.)$. Let $A$ be a linear operator mapping 
$\fron$ into itself. We say that $A$ has a kernel $N$ (with respect to the 
form) when there exists in $\fron \otimes \fron$ a two-point function 
$N (x, y) $ such that  $\forall  \Phi \in \fron $
\beq    \bigl(  A \Phi \bigr)  (y)  = 
       \bigl(   N(y,x) ; \Phi (x)   \bigr)    \label{noy0}        \eeq
where $N_y (x)$ is $\forall y$ an element of $\fron$.
In fact $N$ is a bi-scalar. When $A$ is the identity, we say that $N$ is a 
 {\em reproducing kernel} \cite{aron}   in $\fron$.

\noi  Exemple:    $\fron = \eron _n$,
 eigenspace of the Laplacian in $ (V_2 , \gam )$, for 
a fixed eigenvalue $n$. 
In this example,  the sesquilinear form is  given by (\ref{scalpositif}) and 
  $\fron$ is a Hilbert space.
In  $\eron _n$ the identity admits a  kernel 
\beq      \Gam _n (\xi ,\eta ) =  
 \sum _ {a=1} ^r    E_{a, n} (\xi) \    E_{a, n}  (\eta)     \eeq
We have
 $ \disp   <\Gam (\eta , \xi )  ,  F(\xi ) >  _2  = F (\eta )      $
for all $F \in  \eron _n$, thus $\Gam$ is a  reproducing kernel on 
 $\eron _n$.
In this case,  note that $\Gam (\eta ,\xi )$ actually belongs to $\fron$ as
a function of $\xi$ labelled with $\eta$ (and {\em vice versa}).
Note that  $\Gam$ is real and doesnot depend on the choice of a real 
orthonormal basis in $\eron _n$. It is intrinsically determined by the 
geometry of          $ (V_ 2 , \gam )$ \cite{geomspec}. In this example, 
$\Gam$ is unique because $<.,.>_2  $ is positive definite.

\medskip   \noi
 In fact, requiring that always  $N \in \fron \otimes \fron $   would be too 
restrictive  for the needs of quantum mechanics, therefore the kernels that 
are commonly considered are bi-scalar distributions, say  
$N \in  \fron ' \otimes \fron '  $     where $\fron '$ is a suitable space of 
distributions. 
.

Going back to the problem of positive/negative frequency splitting,   our 
interest is in the possibility for $\Pi ^{\pm}$ to admit a kernel.
The positive/negative-frequency kernel is formally defined as a (continuous) 
linear functional by
\beq  ( \Pi ^{\pm} \psi )(y) =
        \bigl( D^{\pm} (y, x) ;   \psi (x) \bigr)     \label{noy}  \eeq
This functional is actually continuous, because  $\Pi ^{\pm}$ is continuous 
in the sense of some Sobolev-space topology defined on the initial data. 
This point stems from the fact that  (considering  $\psi $ as 
defined by initial data on some Cauchy surface)  $\psi$ varies continuously 
with the initial data (Ref. \cite{wal} p.56) and because the receipe given in 
Appendix 1  doesnot break this continuity.

\medskip \noi
But we proceed mode-wise thus, in each mode $\hron _n$, it is natural
 to consider a kernel for $\Pi ^{\pm} _n$. We take advantage of the 
factorization as follows; 
 let   $D^{\pm} _ {1,n} $  be  a kernel for   $  \Pi ^{\pm}   _{1,n} $   
that is to say,     $\forall f \in  \sron  _n$
\beq      (f^{\pm}) (v)   =  
      \bigl( D^{\pm} _{1,n} (v, u) ;  f (u)  \bigr)_1     \label{noy1}   \eeq
where 
 $ \disp  f^{\pm}   \equiv    \Pi ^{\pm}   _{1,n} \    f   $ by definition.

\medskip
\noi    Now it is not difficult to prove that
\beprop       The bi-scalar
\beq   D^+ _n (y, x)   =  D_{1,n} ^+  (v, u)   \              \ 
                    \Gam _n (\eta , \xi )     \label{canon}                 \eeq
is a kernel  for  $\Pi _n  ^+$. 
It is manifestly invariant under structural isometries.
Since  $  D_{1,n} ^+ (u,v) $ satisfies    {\em  (\ref{reduf}) }
 in  its argument $u$, then
$D^+ _n$  satisfies  {\em  (\ref{weq})}.
\enprop

Proof:

\noi 
Invariance under structural isometries stems from the following observations:
 $\Gam$ is the unique kernel of the identity in $\eron _n$.
 The well-known
 isometric invariance of the Laplacian   $\Del _2$ entails that each
 eigenspace  $\eron _n$ is (globally) invariant  under the isometries of
 $(V_2 , \gam )$.
If $T$ is such an isometry, it leaves invariant the $q$-dimensional scalar
 product  $ <F , G >_2 $.  

\noi
If $F$ is a function on  $ (V_2 , \gam )$  we can write   $T F =  F ( T \xi )$.
Thus $E_{1,n} (u,\xi ) \cdots   E_{r,n} (u,\xi )  $  is another real 
orthogonal basis of  $\eron _n$.  So
$ \Gam _n (T \eta , T \xi )  =  \Gam _n (\eta , \xi )$ and we can write
$$ D_n ^+ (Ty , Tx ) =  D_n ^+ (y , x )  $$
 In other words, any isometry of 
the second factor manifold leaves $D^+ _n $ invariant  [].

\medskip
\noi      Expression (\ref{canon})  will be called {\em canonical}.

\noi   The only arbitrariness involved in formula     (\ref{canon})
  is in the reduced kernel  $D_{1,n} ^+ (v, u)$
  which depends on the choice of a positive-energy projector in $\sron _n$. 

\noi
In general $\sron$ is an infinite dimensional vector space, with the exception 
that $ {\rm dim} \sron = 2$ when ${\rm dim}  V_1 = 1$; this particular case 
has been described in \cite{eres}.

\noi Note that, when $(V_2 , \gam )$ has constant curvature, explicit 
expressions for a basis of  $\eron _n$, hence for $\Gam _n$,
 are available in closed form in the litterature.

\section{Concluding Remarks}
\noi    With help of our  mode decomposition, the problem of finding an 
invariant quantization of free particles in the $p+q$-dimensional warped 
product $V_1 \times V_2$ has been reduced to a similar problem {\em without 
symmetry requirement}, but with a source term, in the $p$-dimensional manifold 
$V_1$ (this reduced problem being one-dimensional, in particular, 
 when we start from a generalized FRW spacetime).

\noi We have characterized a family of admissible complex-structure operators.
Each one  uniquely corresponds to a splitting of the solutions to the 
primitive problem into
 positive-frequency   and negative-frequency parts. Our procedure respects 
isometric invariance, at least insofar as all the 
isometries of $V_1 \times V_2$ are induced by symmetries of its  second factor 
 (structural isometries).
This case already encompasses   a very large class of spacetimes.

\noi When  there are isometries induced by symmetries in the first factor 
manifold, the situation is still partially under control. For instance, the 
case where $(V_1 , \alp )$ is stationary can be handled, and gives rise to an 
operator $J$ which commutes not only with structural isometries, but also with 
the time translations.

           \noi In contrast, there is no clue for the case where an 
extraordinary Killing field exists. Fortunately, the occurence of this case is 
limited by the  severe condition (\ref{severe}) 
involving the warping factor $\sqrt{S}$.

\noi Note that our approach is concerned  with one given structure of warped 
spacetime; it would become ambiguous in the degenerate cases where $V$ can be 
regarded as warped in more than one manner. This remark applies to de Sitter 
space; fortunately, in that case, it is possible   to construct an invariant 
vacuum by a different method \cite{sit}.

\noi In this paper we have considered spacetimes with smooth metrics; 
extension to more realistic situations requires further work.

The most  general question as to know under which conditions  the free motion 
of scalar particles in  an "arbitrary spacetime" bearing Killing vectors 
admits an {\em isometrically invariant}  quantization, remains  open.
However, isometric invariance is more easily  implemented within the 
framework of warped spacetimes.

$$   \  $$
$$ \    $$
{\bf APPENDIX 1}

\noi {\bf Klein-Gordon equation with a "source term"}

\noi   Consider $N$ dimensional spacetime  $V_N$, with coordinates 
$x^0,  x^1 , \cdots  x^{N-1} $.

\noi   Consider the KG equation with a nonderivative external coupling
\beq     \nab ^2   \Phi     +   A (x)   \Phi   = 0  
           \label{coupnonderiv}   \eeq
where $A$ is a smooth function.

\noi  For {\em real} solutions  $\Phi , \Omega $ the vector field
  $   \Phi \nab ^\alp \Omega  -   \Omega \nab ^\alp \Phi  $ 
 is conserved.  Under suitable  technical assumptions 
  the quantity
\beq \varpi  ( \Phi , \Omega ) =    \int _\Sig    j \cdot d \Sig =
\int _ \Sig  (\Phi \nab ^\alp \Omega - \Omega \nab \Phi ) \   d \Sig _ \alp   
                                   \eeq                                       
is finite and doesnot depend on the choice of the spacelike hypersurface 
$\Sig$. It defines a skew-symmetric form on the linear  space of solutions to 
(\ref{coupnonderiv}).
              
\noi 
When    $\Sig$ is defined by  $x^0 = 0$ we have {\em on this hypersurface}
$ \disp    j \cdot \Sig = j ^0 \   d \Sig _ 0 $ hence
$$     d \Sig _ 0 =  \sqrt {|g|} \   d^{N-1}  x     $$

\noi     If we can choose coordinates such that 
 $ \disp   \       g ^{0 \alp } | _ \Sig   = 0  \qquad
 {\rm  for} \qquad \alp \not = 0 \      $
we can simply write 
\beq        \varpi ( \Phi , \Omega ) =      \int _ {x^0 = 0} 
  (\Phi \partial _0  \Omega - \Omega \partial _ 0  \Phi ) \ 
g ^{00}   \     \sqrt{|g|} d ^{N-1} x            \label{symp.N}         \eeq
Let us stress that $\varpi$ is intrinsically defined.
In contradistinction,  a complex structure operator $J$ acting on 
the solutions  to  (\ref{coupnonderiv})  is by no means unique.
Such a $J$   can  be  associated to each Cauchy surface.
The receipe proposed below (valid in the presence of a source term) 
is inspired from, but not identical with,  the procedure indicated by 
Ashtekar and Magnon  in the context of  minimal coupling. 
In addition, we shall apply this receipe, not in  the full spacetime
 $ (V, g)$, but within its first factor manifold, $V_1$. 

\noi  In order to build a complex structure  we remind  that each  solution of 
(\ref{coupnonderiv})  is  uniquely and globally determined  by its value and 
that of its time derivative on any Cauchy surface $\Sig$,
 see p.56 of reference  \cite{wal}.

\noi Therefore, the space of solutions to (\ref{coupnonderiv})
is isomorphic to  the vector space $\cron _\Sig$  of couples
   \[  \left(   \begin{array}{c}  U    \\   V    \end{array}  \right)  \]  
 where $U$ and $V$ are smooth functions on the Cauchy surface  $\Sig$,  
each solution $\Phi$ being  represented by the couple
\[   \left(    \begin{array}{c}  \Phi |_\Sig    \\
      (\partial _ 0  \Phi )_{\Sig}     \end{array}      \right)     \] 
Naturally  $\varpi$  induces a skew form on these couples,
 denoted by the same typographic character, say  
\beq     \varpi \quad   \bigl(
   \left(   \begin{array}{c}  U    \\   V    \end{array}  \right)    
\  ,                                               
     \left(   \begin{array}{c}  U'    \\   V'    \end{array}  \right)     
\bigr)            =
\int    (U V' -  U' V ) \     g^{00}  \sqrt{|g|} \    d^{N-1}   x     
                                     \label{skewcoupl}                   \eeq

Let $J$ be defined by
\beq  J   \left(   \begin{array}{c}  U    \\   V    \end{array}  \right)  =
 \left(    \begin{array}{c}   -  V \\    U      \end{array}      \right) 
                                       \label{defJ}      \eeq
   Of course, $J$  depends on $\Sig$.  According to our assumptions,
 it is clear that the new   Cauchy data
$$ (J \Phi )_ \Sig  = -   V  , \qquad  \quad
       ( \partial _0 (J \Phi ) ) _\Sig =   U    $$ 
globally  define  $J \Phi$  as another  solution.

\noi   Then, using  expression  (\ref{symp.N}) or  (\ref{skewcoupl})
 it is not difficult to check that

1) $J^2 = -1$

2) $J$ leaves  $\varpi$ invariant

3) $   \varpi(\Phi , J \Phi ) > 0 $  when  $    \Phi \not= 0 $.

4)  
Moreover, as $\Phi $ varies continuously with the initial data (Wald p.56) $J$ 
is a continuous operator in the sense of a suitable Sobolev topology.

We have this result

\beprop             Let $T$ be an isometry  of $(V, g)$. 
If  $T$   leaves invariant   the Cauchy surface  $\Sig$ and 
 the source term (that is  $A(Tx) = A(x)$), 
  then  the complex-structure operator $J$  defined as above is 
invariant by $T$.
\enprop

Proof:
In view of  the above assumptions,
not only  $T$ induces a transformation of the functions $\phi (x)$, but also a 
transformation of the functions $U, V$  defined on $\Sig$, say with a slight 
abuse of notation,
$ (T U) (y) = U (Ty)$ where $y \in \Sig$. And $T$ is natural with respect to 
restrictions to $\Sig$, that is
$\disp  (T \phi ) |_{\Sig}   =   T  \bigl(   \phi  |_{\Sig}     \bigr)  $.
Represent  $\phi$ by  
$ \disp    \left(   \begin{array}{c}  U    \\   V    \end{array}  \right) $
and    $J$ as in  (\ref{defJ}). We obtain       $ [J, T ]=0$.

\bigskip
    \noi
{\bf APPENDIX 2 

\noi  The Complex Structure}       

\noi Consider a real vector space $\calk$  endowed with a skew-symmetric form 
$w$. By complexification we obtain  $\calk  ^C$ and extend $w$ to it. 
It turns out that   $\calk  ^C$ is  endowed with a sesquilinear form
\beq  (\Phi ; \Psi)   =   -i  w (\Phi ^*  ,  \Psi)   \label{sesq2}   \eeq
           for      $    \Phi   ,  \Psi   \in  \calk ^C $.
We say that $\Phi$ and $\Psi$ are mutually {\em orthogonal} when 
$ (\Phi ; \Psi)$  vanishes.

\noi  Let $B$ be a linear  operator acting in    $\calk  ^C$.
 We say that $B$ is {\em symmetric}  with respect to the sesquilinear form when
$       ( B \Phi ; \Psi)   = (\Phi ; B  \Psi ) $
for all $\Phi, \Psi$. Just as well as in  Hilbert spaces, symmetric operators 
in this sense have orthogonal eigenspaces (the proof is straightforward, but 
it may happen that $ (\phi ; \phi ) $ vanishes for some $\phi \not= 0$.

\medskip \noi
 A {\em complex-structure operator}  on  $\calk$ is a  {\em real}
linear  operator $J$  such that  $J^2 = -1$. 
Its extension to $\calk ^C$  is real in this sense that
   $ (J \Psi) ^*  =   J  \Psi ^* $.

\noi
Let  $ \disp  \Pi ^\pm = \half (1 \pm i J)$. Note that $\Pi ^\pm$ are not 
real, indeed   $(\Pi ^+  \Psi )^* =  \Pi ^-   \Psi ^* $.

\noi "Positive-frequency" vectors can be characterized equivalently by either
$\Pi^+ \Psi = \Psi$  or     $ J \Psi =  -i \Psi$.

\noi        It is easy to check 
that, if $J$ is  positive with respect to the skew-symmetric form, 
then    $\Pi ^+ \Psi  = \Psi $   implies that 
 $ (\Psi ; \Psi ) > 0   \qquad  \      \forall  \Psi \not= 0 $ 
 (resp.   $  \Pi ^- \Psi  = \Psi , \qquad   (\Psi ; \Psi ) < 0 $).

Proof

Let  $ \Psi =  M + i N $ with $M$ and $N$ real, then we have
\beq   (\Psi ; \Psi ) \equiv 2 w (M, N)      \label{PsiPsi}    \eeq 
 which is real anyway.
If now we assume that $ \Pi ^+ \Psi = \Psi$, on the one hand we obtain
\beq  J \Psi =  - i  \Psi                 \eeq 
  where   $ - i \Psi =  N -  i M   $.
On  the other hand   we have $J \Psi = J M + i J N    $. 
 By identifying we obtain
\beq    J M = N , \qquad   J N = - M           \eeq
Thus in  (\ref{PsiPsi})    we can replace $N$    by  $ JM$.
Now we have   
$$     (\Psi ; \Psi ) =  2 w (M, J M)      $$
But positivity of $J$  ensures that   $ w(M, JM)$ is positive,
 unless $M$ vanishes.
The case where $M$ vanishes is necessarily that where $\Psi$ vanishes,
 for $ N=JM$. 

In addition, invariance of $ w$ by  $ J $  ensures that $\Pi ^\pm$ is 
symmetric with respect to the sesquilinear form. 

Proof:
Observe that (owing to $J^2 = -J$) the complex-structure operator is skew-
symmetric with respect to the skew form,  that is 
\beq    w  (\Phi , J \Psi ) =  -  w (J \Phi , \Psi ) 
  \qquad   \qquad  \forall \Phi , \Psi       \eeq
iff    $ w  $  is invariant by $J$.         Then,  it is 
straightforward to check that  this property entails the skew-symmetry
 of $J$ with respect to the sesquilinear form, namely
$$  ( J \Phi ; \Psi ) =  -i w ( J \Phi ^*  ,  \Psi ) =
     i w ( \Phi ^* , J  \Psi )  =  -  ( \Phi ; J \Psi )    $$
But since  $\Pi ^{\pm}  = \half (1 \pm   iJ)$   it follows that 
               $\Pi ^{\pm} $ is symmetric with respect to (\ref{sesq2}).


\begin{thebibliography}{3}
\bibitem{wal} R. M. Wald {\em Quantum Field Theory in Curved Spacetime and 
Black Hole Thermodynamics} Chicago Lectures in Physics, The University of 
Chicago Press, Chicago (1994). See especially pp. 96-97.                   

\bibitem{cheval} M. Chevalier, J. Math. Pures et Appl. {\bf 53}, 223 (1974);
see also E. Combet, S\'eminaire Phys. Math. Coll\`ege de France (1965).

\bibitem{mor} C. Moreno, J. Math. Phys. {\bf 18}, 2153 (1977)



\bibitem{ashmag} A. Ashtekar and A. Magnon 
Proc. Roy. Soc. London, {\bf A 346}, 375 (1975). In that  paper, aiming at the  
 development of the algebraic approach, kernels are not explicitly  considered.

  

\bibitem{moreno} C. Moreno 
Reports in Theoretical Physics, 333-358 (1980)

\bibitem{lichhouch} A. Lichnerowicz, in
 {\em Relativity, Groups and Topology}, Les Houches 1963,
De Witt and De Witt eds. Science Publishers Inc. New York 1964

\bibitem{eres}  Ph. Droz-Vincent,
Isometric invariance of the positive-frequency kernel in generalized FRW 
spacetimes, contribution to ERES 98, Salamanca, 
{\em in Relativity and Gravitation in General},
Editors J. Martin, E. Ruiz, F. Atrio and A. Molina,
World Scientific Publishing Co, Singapore, London, Hong-Kong (1999).


\bibitem{lichne} A. Lichnerowicz, {\em Propagateurs et Commutateurs en 
Relativit\'e G\'en\'e rale}, Inst. des Hautes Etudes Sci. 
Publications Math\'ematiques $n^0 $ 10, Section 6, p. 16.   



\bibitem{sit} N.A. Chernikov and E.A. Tagirov, 
  Ann.  Inst.H.Poincar\'e, {\bf 9A}, 109 (1968).
 these authors  mainly consider {\it nonminimal coupling\/}.
  In addition see also:  
   J. Geheniau and  Ch. Schomblond,
 Acad. R. de Belgique, Bull.  Cl. des Sciences  {\bf 54},  1147  (1968).
 Ch. Schomblond and  Ph. Spindel,
 Ann. Inst. H. Poincar\'e, {\bf 25},          67     (1976)
  (these works are  concerned with 
the "steady-state" manifold rather than global De Sitter). 

\noi  B. Allen,  Phys. Rev. D {\bf 32}, 3136 (1985).


\bibitem{bir}  Birell  and  Davies, 
{\em Quantum Fields in Curved Space} (Cambridge Univ. 
Press 1982) Chap.3, p. 58  for a statement of invariance of the Wightman 
"function" in an asymptotically static, spatially flat,  FRW universe.  

\bibitem{carot}  Jaime Carot and J. da Costa
On the geometry of warped spacetimes,
{\em Class. Quantum. Grav.} {\bf 10},   461-482 (1993)


\bibitem{sanch}  M. S\'anchez,
On the geometry of generalized Robertson-Walker spacetimes: curvature and 
Killing fields,  
 J. Geom. and Phys. {\bf 31}, 1-15 (1998).





\bibitem{geomspec}             M.Berger, P.Gauduchon, E.Mazet 
   " Lecture Notes in Math." {\bf 194}
                  Springer Verlag Berlin, Heidelberg, N-York    1971.
 S. Gallot, D. Hulin, J. Lafontaine,
"Riemannian Geometry"  Springer Verlag, Berlin, Heidelberg, N-York   1987      
                       Chap. IV D, pp. 196-202



\bibitem{segal} I. Segal, J. Math. Phys.{\bf 1}, 468 (1960).

\bibitem{ellhawk}  S.W. Hawking and  G.F.R. Ellis,
{\em The large scale structure of space-time}, Cambridge Univ. Press 
(Cambridge 1973) Chap. 7,  Section 7.4,  pp 233-243.

\bibitem{droz}    Ph. Droz-Vincent, Clas. Quant. Grav.{\bf 17}, 1-17  (2001).
In order to avoid typographic confusion with new quantities introduced now, we 
make a few notational changes with respect to  that article:

\noi
Replacement   of $D$ by   $\Del _1 ^\sharp$   Replacement  of $J^A$ by $Y^A$.

                                   
 \bibitem{aron}    Strictly speaking, the notion of a  reproducing kernel
 should be limited to operators on  Hilbert spaces.


\end{thebibliography}
\end{document}